\documentclass{amsproc}

\newcommand{\HH}{\mathcal{H}}
\newcommand{\CC}{\mathbb{C}}
\newcommand{\RR}{\mathbb{R}}
\newcommand{\ZZ}{\mathbb{Z}}

\newcommand{\LL}{\mathcal{L}}

\newcommand{\QQ}{\mathbb{Q}}

\newcommand{\TT}{\mathbb{T}}
\newcommand{\SSS}{\mathbb{S}}

\newcommand{\Arg}{\mathop{\mathrm{Arg}}}
\newcommand{\dom}{\mathop{\mathrm{dom}}}
\newcommand{\ran}{\mathop{\mathrm{ran}}}
\newcommand{\spec}{\mathop{\mathrm{spec}}}
\newcommand{\diag}{\mathop{\mathrm{diag}}}

\newtheorem{theorem}{Theorem}
\newtheorem{prop}[theorem]{Proposition}
\newtheorem{lemma}[theorem]{Lemma}

\theoremstyle{definition}

\begin{document}

\author{Konstantin Pankrashkin}

\address{L.A.G.A., Institut Galil\'ee, Universit\'e Paris-Nord,
99 Jean-Baptiste Cl\'ement, 93430 Villetaneuse, France \&
Institut f\"ur Mathematik, Humboldt-Universit\"at zu Berlin, Rudower
Chaussee 25, 12489 Berlin, Germany}

\email{const@math.hu-berlin.de}

\title{Localization in a quasiperiodic model on quantum graphs}

\dedicatory{Dedicated to the memory of Vladimir Geyler (1943--2007)}

\begin{abstract}
We show the presence of a dense pure point spectrum on quantum graphs
with Maryland-type quasiperiodic Kirchhoff coupling constants at the vertices.
\end{abstract}

\maketitle

\section*{Introduction}

In the present contribution we are going to show how quantum graphs
can be used to construct exactly solvable quasiperiodic models
showing the Anderson localization at all energies. 

In~\cite{GFP} it was shown that the one-dimensional
quasiperiodic difference Hamiltonian
\begin{equation}
 \label{eq-ml}
L\psi(n)=\psi(n+1)+\psi(n-1)+\lambda\tan(\omega n-\alpha)\psi(n),
\quad \psi\in \ell^2(\ZZ),\quad \lambda,\alpha,\omega\in\RR,
\end{equation}
has a pure point spectrum dense everywhere under some arithmetic
conditions for $\omega$ and $\alpha$; this operator
is often referred to as the Maryland model.
Later the class of such Hamiltonians was considerably extended
in several directions, e.g. to the multidimensional case
with more general coefficients~\cite{B1,FP}, see also~\cite{BBP}
for a recent review. It is a rather interesting problem
to construct continuous quasiperiodic operators
where one can describe the dense point spectrum in a more
or less explicit way. An example of such models was proposed in~\cite{gm},
where Schr\"odinger operators with tan-like quasiperiodic
point perturbations were studied and the Anderson localization
in the gaps of the unperturbed Hamiltonians was shown.
In the paper~\cite{exdual} a comb-shaped quantum graph was proposed
whose spectral study reduces at the formal level
to operators of the form~\eqref{eq-ml}; however,
the machinery used does not allow to prove rigorously the presence
of a dense pure point spectrum in that case.
We are going to show, using some modification of the constructions
from~\cite{FP} and~\cite{gm} and the machinery of self-adjoint extensions~\cite{BGP},
that the Anderson localization at all energies
can be achieved by placing tan-like quasiperiodic $\delta$-interactions
at the vertices of quantum graph lattices. We consider this case
as the most illustrative one in many aspects, and we plan to
treat much more general quasiperiodic interactions in subsequent works.

\section{Schr\"odinger operator on a quantum graph}

Below we describe some basic constructions for quantum graphs; a detailed discussion
can be found e.g. in~\cite{GS,Ku1,Ku2}. There are many approaches to the study
of the spectra of quantum graphs, we use the one from~\cite{BGP,KP1} based
on the theory of self-adjoint extensions.

We consider a quantum graph whose set of vertices is identified with $\ZZ^d$, $d\ge 1$.
By $h_j$, $j=1,\dots,d$, we denote the standard basis vectors of $\ZZ^d$.
Two vertices $m$, $m'$ are connected by an oriented edge $m\to m'$ iff $m'=m+h_j$
for some $j\in\{1,\dots,d\}$; this edge is denoted as $(m,j)$
and one says that
$m$ is the initial vertex and $m'\equiv m+h_j$ is the terminal vertex.

Fix some $l_j>0$, $j\in\{1,\dots,d\}$, and
replace each edge $(m,j)$ by a copy of the segment $[0,l_j]$
in such a way that $0$ is identified with $m$ and $l_j$ is identified with $m+h_j$.
In this way we arrive at a certain topological set carrying a natural metric structure. 

The quantum state space of the system is 
\[
\HH:=\bigoplus_{(m,j)\in\ZZ^d\times\{1,\dots,d\}} \HH_{m,j},\quad
\HH_{m,j}=\LL^2[0,l_j],
\]
and vectors $f\in\HH$ will be denoted as $f=(f_{m,j})$, $f_{m,j}\in\HH_{m,j}$,
$m\in\ZZ^d$, $j=1,\dots,d$.
Let us describe the Schr\"odinger operator acting in $\HH$. Fix some real-valued
functions (potentials) $U_j\in \LL^2[0,l_j]$, $j=1,\dots,d$, and some real constants
$\alpha(m)$, $m\in\ZZ^d$. Set $A:=\diag\big(\alpha(m)\big)$; this is a self-adjoint operator
in $\ell^2(\ZZ^d)$. Denote by $H_A$ an operator
acting as
\begin{subequations}
      \label{eq-sch}
\begin{equation}
        \label{eq-act}
(f_{m,j})\mapsto \Big(
(-D^2+U_j)f_{m,j}\Big),\quad Df_{m,j}=f'_{m,j},
\end{equation}
on functions $f=(f_{m,j})\in\bigoplus_{m,j} H^2[0,l_j]$
satisfying the following boundary conditions:
\begin{equation}
      \label{eq-cont1}
f_{m,j}(0)=f_{m-h_k,k}(l_k)=:f(m),\quad j,k=1,\dots,d,\quad m\in\ZZ^d,
\end{equation}
(which means the continuity at all vertices)
and
\begin{equation}
f'(m)=\alpha(m)f(m),\quad m\in\ZZ^d,
\end{equation}
where
\begin{equation}
f'(m):= \sum_{j=1}^d f'_{m,j}(0)-\sum_{j=1}^d f'_{m-h_j,j}(l_j).
\end{equation}
\end{subequations}
The constants $\alpha(m)$ are usually referred to as \emph{Kirchhoff coupling constants}
and interpreted as the strengths of zero-range impurity potentials at the corresponding vertices~\cite{exlat}.
The zero coupling constants correspond hence to the ideal couplings
and are usually referred to as the standard boundary conditions.

Denote by $S$ the operator acting as \eqref{eq-act} on the functions $f$
satisfying only the boundary conditions \eqref{eq-cont1}. On the domain
of $S$ one can define linear maps
\[
f\mapsto \Gamma f:= \big(f(m)\big)_{m\in\ZZ^d}\in \ell^2(\ZZ^d),\quad
f\mapsto \Gamma' f:= \big(f'(m)\big)_{m\in\ZZ^d}\in \ell^2(\ZZ^d).
\]
By the Sobolev embedding theorems, $\Gamma,\Gamma'$ are well-defined,
and the joint map $(\Gamma,\Gamma'):\dom S\to \ell^2(\ZZ^d)\times \ell^2(\ZZ^d)$ is surjective.
Moreover, by a simple algebra, for any $f,g\in\dom S$ one has
$\langle f,Sg\rangle-\langle Sf,g\rangle=\langle \Gamma f,\Gamma' g\rangle-
\langle\Gamma' f,\Gamma g\rangle$ (see e.g. proposition~1 in~\cite{KP1}).
In the abstract language, $(\ZZ^d,\Gamma,\Gamma')$ form a \emph{boundary triple}
for $S$. This permits to write a useful formula for the resolvent of $H_A$,
which will play a crucial role below.

First, denote by $H^0$ the restriction of $S$ to $\ker \Gamma$.
Clearly, $H^0$ acts as \eqref{eq-act}
on functions $(f_{m,j})$ with $f_{m,j}\in H^2[0,l_j]$
satisfying the Dirichlet boundary conditions,
$f_{m,j}(0)=f_{m,j}(l_j)=0$ for all $m,j$, and the spectrum
of $H^0$ is just the union of the Dirichlet spectra of the operators
$-\dfrac{d^2}{dt^2}+U_j$ on the segments $[0,l_j]$.
We will refer to $\spec H^0$ as to the \emph{Dirichlet spectrum}
of the graph.

Denote by $s_j$ and $c_j$ the solutions to
$-y''+U_j y=zy$ satisfying $s_j(0;z)=c'_j(0;z)=0$
and $s'_j(0;z)=c_j(0;z)=1$, $z\in\CC$, $j=1,\dots,d$.

For $z$ outside $\spec H^0$ consider operators $\gamma(z):\ell^2(\ZZ^d)\to \HH$ defined as follows.
For $\xi\in \ell^2(\ZZ^d)$, $\gamma(z)\xi$ is the unique solution to $(S-z)f=0$ with $\Gamma f=\xi$.
For each $z$, $\gamma(z)$ is a linear topological isomorphism
between $\ell^2(\ZZ^d)$ and $\ker(S-z)$. 
Clearly, in terms of the functions $s_j$ and $c_j$ introduced above,
one has
\begin{multline*}
\big(\gamma(z)\xi\big)_{m,j}(t)=
\dfrac{1}{s_j(l_j;z)}\Big(\xi(m+h_j)s_j(t;z)\\
+\xi(m)\big(
s_j(l_j;z) c_j(t;z)-c_j(l_j;z)s_j(t;z)
\big)\Big),\\
t\in[0,l_j],\quad (m,j)\in\ZZ^d\times\{1,\dots,d\}.
\end{multline*}
Furthermore, for the same $z$'s define an operator $M(z):\ell^2(\ZZ^d)\to \ell^2(\ZZ^d)$
by $M(z):=\Gamma'\gamma(z)$. In our case,
\[ M(z)\xi(m)=
\sum_{j=1}^d \dfrac{1}{s_j(l_j;z)}\cdot\Big( \xi(m-h_j)+\xi(m+h_j)
-\eta_j(z)\,\xi(m)\Big),\quad \xi\in\ell^2(\ZZ^d),
\]
where $\eta_j(z):=c_j(l_j;z)+s'_j(l_j;z)$ is the Hill discriminant associated with $U_j$.
The maps $\gamma$ and $M$ satisfy a number of important properties.
In particular, $\gamma$ and $M$ depend analytically on their argument (outside of $\spec H^0$),
$M(z)$ is self-adjoint for real $z$,
\begin{gather}
        \label{eq-meg}
\text{for any non-real $z$ there is $c_z>0$ with}
\dfrac{\Im M(z)}{\Im z}\ge c_z, \text{ and}\\
        \label{eq-meg2}
M'(\lambda)=\gamma^*(\lambda)\gamma(\lambda)>0 \text{ for } \lambda\in\RR\setminus\spec H^0.
\end{gather}
Furthermore,
\begin{equation}\label{eq-nz}
\gamma^*(z) f=0 \text{ for any }f\in\ker(S-z)^\perp\equiv\gamma(z)\big(\ell^2(\ZZ^d)\big)^\perp,
\end{equation}
see \cite[Section~1]{BGP}.

\begin{prop}\label{prop-krein}
The resolvents of $H^0$ and $H_A$ are related by the Krein resolvent formula, 
\begin{equation}
          \label{eq-krein}
(H_A-z)^{-1}=(H^0-z)^{-1}-\gamma(z)\big(M(z)-A\big)^{-1}\gamma^*(\bar z),
\quad z\notin\spec H_A\cup \spec H^0,
\end{equation}
and the set $\spec H_A\setminus \spec H^0$ coincides with
$\{z\notin\spec H^0:\, 0\in\spec \big(M(z)-A\big)\}$.
For any $z\notin\spec H^0$ there holds $\ker(H_A-z)=\gamma(z)\ker\big(M(z)-A\big)$,
i.e. $z$ is an eigenvalue of $H_A$ iff $0$ is an eigenvalue of $M(z)-A$,
and $\gamma(z)$ is an isomorphism of the corresponding eigensubspaces.
\end{prop}

\section{Eigenvalues for Maryland-type coupling constants}

We are going to study the above operator $H_A$ for a special choice of
the coefficients $\alpha(m)$. Namely, pick $g>0$, $\omega\in\RR^d$,
$\varphi\in\RR$ with 
\begin{equation}
   \label{eq-vf1}
\varphi\ne \pi/2-\pi\langle\omega, m\rangle \mod \pi
\end{equation}
and set
\[
\alpha(m):=-g\tan\big(\pi\langle \omega, m\rangle+\varphi\big),\quad m\in\ZZ^d.
\]
We assume additionally that $\omega$ satisfies the following \emph{Diophantine
condition}:
\begin{equation}
\label{eq-vf2}
\text{there are }C,\beta>0 \text{ with }|\langle \omega, m\rangle-r|\ge C|m|^{-\beta}
\text{ for all }m\in\ZZ^d\setminus\{0\},\, r\in\ZZ.
\end{equation}
Clearly, \eqref{eq-vf2} implies $\omega\notin\QQ^d$.

The operator $H_A$ corresponding to the above choice of the coupling
constants will be denoted simply by $H$, and this will be our main object of study.
Our main result concerning the spectrum of $H$ is the following theorem.

\begin{theorem}\label{th-main}
The spectrum of $H$ is pure point and coincides with $\RR$.
\end{theorem}

We will repeat first some algebraic manipulations in the spirit of~\cite{FP}.
Let $U$ be the multiplication by the sequence $\big(e^{2\pi i\langle\omega,m\rangle}\big)$ in $\ell^2(\ZZ^d)$,
$B(z):=\big(M(z)-ig\big)^{-1}$, $C(z):=-\big(M(z)+ig\big)\big(M(z)-ig\big)^{-1}$.
The operators $B(z)$ and $C(z)$ are defined at least for $z$ with $\Re z\notin\spec H^0$
and $|\Im z|$ sufficiently small.
Denoting $\chi:=e^{2i\varphi}$ one can write for such $z$ the identity
\begin{equation}
     \label{eq-MA}
M(z)-A=B(z)^{-1}\big(1-\chi C(z)U\big)(1+\chi U)^{-1} .
\end{equation}

In what follows we denote $\SSS^1:=\{z\in\CC,\,|z|=1\}$ and $\TT^d:=\underbrace{\SSS^1\times\dots\times\SSS^1}_{d \text{ times}}\subset\CC^d$. For $\theta=(\theta_1,\dots,\theta_d)\subset\CC^d$
and $m=(m_1,\dots,m_d)\in\ZZ^d$ we write $\theta^m:=\theta_1^{m_1}\dots\theta_d^{m_d}$,
and in this context $k\in\ZZ$ will be identified with the vector $(k,\dots,k)\in\ZZ^d$,
i.e. $\theta^{-1}:=\theta_1^{-1}\dots\theta_d^{-1}$ etc. 

Denote by $F$ the Fourier transform carrying $\ell^2(\ZZ^d)$ to $\LL^2(\TT^d)$,
\[
F\psi(\theta)=\sum_{m\in\ZZ^d} \psi(m) \theta^m,\quad
F^{-1}f(m)=\dfrac{1}{(2\pi i)^d} \int_{\TT^d} f(\theta) \theta^{-m-1}d\theta.
\]
Under this transformation $M(z)$ becomes the multiplication by a function
$M(z,\theta)$,
\begin{equation}
   \label{eq-man}
M(z,\theta):=
\sum_{j=1}^d\dfrac{1}{s_j(l_j;E)} \, (\theta_j+\theta_j^{-1})
- 
\sum_{j=1}^d \dfrac{\eta_j(E)}{s_j(l_j;E)},
\end{equation}
the operators $B(z)$ and $C(z)$ become the multiplications
by $B(z,\theta):=\big(M(z,\theta)-ig\big)^{-1}$
by $C(z,\theta):=-\big(M(z,\theta)+ig\big)\big(M(z,\theta)-ig\big)^{-1}$,
respectively, and $U$ becomes a shift operator,
$U k(\theta_1,\dots,\theta_d)=k(e^{2\pi i \omega_1}\theta_1,\dots,e^{2\pi i \omega_d}\theta_d)$.

Consider an arbitrary segment $[a,b]\subset\RR\setminus\spec H^0$.
Eq.~\eqref{eq-meg}, the analyticity of $\gamma$, and the self-adjointness
of $M(z)$ for real $z$ imply the existence of $\delta'>0$
such that $\|\Im M(z)\|\le g/2$ for $z\in Z:=\{z\in\CC:\, |\Im z|\le\delta',\ \Re z\in[a,b]\}$.
At the same time, this means that $|\Im M(z,\theta)|\le g/2$ for $z\in Z$.
As follows from \eqref{eq-man}, $M(z,\theta)$ can be continued to an analytic function
in $Z\times \Theta$, $\Theta:=\{(\theta_1,\dots,\theta_d)\subset \CC^d:, r<|\theta_j|<R\}$, $0<r<1<R<\infty$.
Choosing $r$ and $R$ sufficiently close to $1$ one immediately sees
that the function
\[
C(z,\theta):=\dfrac{g^2-\big(\Im M(z,\theta)\big)^2-\big(\Re M(z,\theta)\big)^2-2ig\Re M(z,\theta)}{|M(z,\theta)-ig\big|^2}
\]
does not take values in $(-\infty,0)$ for $(z,\theta)\in Z\times\Theta$.
Therefore, the function $f(z,\theta):=\log C(z,\theta)$ is well-defined
and analytic in $Z\times\Theta$,
where $\log$ denotes the principal branch of the logarithm.
We will use the following assertion \cite[Lemma 3.2]{FP} implied by the Diophantine property~\eqref{eq-vf2}:
\begin{lemma}\label{lem1}
The operator $1-U$ is a bijection on the set of functions $v$ analytic in $\Theta$
with $\int_{\TT^d} v(\theta)\theta^{-1}d\theta=0$.
\end{lemma}
By lemma~\ref{lem1}, the function $t(z,\theta):=(1-U)^{-1}\big(f(z,\theta)-f_0(z)\big)$
is well-defined and analytic in $Z\times\Theta$, where
\begin{equation}
     \label{eq-f0}
f_0(z):=
\dfrac{1}{(2\pi i)^d}\int_{\TT^d} f(z,\theta)\theta^{-1}d\theta.
\end{equation}

\begin{lemma}\label{prop-1}
The function $f_0$ is analytic in $Z$,
\begin{gather}
   \label{eq-f01}
\Re f_0(z)<0 \quad \text{for} \quad\Im z>0,\\
   \label{eq-f02}
\Re f(z,\theta)=\Re t(z,\theta)=\Re f_0(z)=0 \quad \text{for} \quad \Im z=0.
\end{gather}
For real $\lambda$ one has $f_0(\lambda)=2i \sigma(\lambda)$, where
\[
\sigma(\lambda)=\dfrac{1}{(2\pi i)^d}\int_{\TT^d} \arctan \dfrac{M(\lambda,\theta)}{g}\,\,\theta^{-1}d\theta.
\]
The function $\sigma$ is real-valued,
strictly increasing, and continuously differentiable on $[a,b]$.
\end{lemma}

\begin{proof}
The analyticity of $f_0$ follows from its integral representation.
Eq.~\eqref{eq-f01} follows from \eqref{eq-f0} if one takes into account
the inequalities $\Im M(z,\theta)>0$ for $\Im z>0$ and $\Re\log z<0$
for $|z|<1$. Equalities~\eqref{eq-f01} follows from from \eqref{eq-f0}
and the real-valuedness of $M(z,\theta)$ for real $z$.

By elementary calculations, for $x\in\RR$ and $y>0$ one  has
\begin{equation}
          \label{eq-hh2}
g_1(x):=\dfrac{1}{2i}\log \dfrac{iy+x}{iy-x}\equiv \arctan \dfrac{x}{y}=:g_2(x).
\end{equation}
In fact, this follows from 
\begin{equation}
    \label{eq-hh}
g'_1(x)=g'_2(x)= \dfrac{y}{x^2+y^2}
\end{equation}
and $g_1(0)=g_2(0)=0$.
Eq.~\eqref{eq-hh2} obviously implies $f_0(\lambda)=2i\sigma(\lambda)$ for $\lambda\in\RR$.
Furthermore, as follows from \eqref{eq-hh},
\[
\sigma'(\lambda)= \dfrac{1}{(2\pi i)^d}\int_{\TT^d} \dfrac{g M'_\lambda(\lambda,\theta)}{M(\lambda,\theta)^2+g^2} \,\,\theta^{-1}d\theta,
\]
and, by \eqref{eq-meg2}, $\sigma'(\lambda)>0$.
\end{proof}

An immediate corollary of the analyticity of $f_0$ and of \eqref{eq-f01} is
\begin{lemma}\label{lem-eps}
There exists $\varepsilon_0>0$ such that $\big|e^{f_0(\lambda)}\xi-1\big|\le 2\big|e^{f_0(\lambda+i\varepsilon)}\xi-1\big|$
for all $\xi\in\SSS^1$, $\lambda\in[a,b]$, and $\varepsilon\in[0,\varepsilon_0]$.
\end{lemma}

Denote by $t(z)$ and $f(z)$ the multiplication operators by $t(z,\theta)$ and $f(z,\theta)$
in $\LL^2(\TT^d)$, respectively.
By definition of $t(z,\theta)$ for any $\varphi\in\LL^2(\TT^d)$
\begin{multline}
        \label{eq-comm}
 e^{t(z)} e^{f_0(z)} U e^{-t(z)}\varphi(\theta)\\
 =
  e^{t(z,\theta)} e^{f_0(z,\theta)} \exp\big({}-t(z,e^{2\pi i \omega_1}\theta_1,e^{2\pi i \omega_d}\theta_d)\big) U \varphi(\theta)\\
  =\exp\big(t(z,\theta)-Ut(z,\theta)+f_0(z,\theta)\big) U\varphi(\theta)=e^{f(z)}U\varphi(\theta)=C(z)U\varphi(\theta). 
\end{multline}
Therefore, one can rewrite Eq.~\eqref{eq-MA} as
\begin{equation}
     \label{eq-MA2}
M(z)-A=B(z)^{-1} e^{t(z)}\big(1-e^{f_0(z)}\chi U\big)e^{-t(z)}\big(1+\chi U\big)^{-1}.
\end{equation}

\begin{prop}\label{prop-ev} The set of the eigenvalues of $H$ in $[a,b]$
is dense and 
coincides with the set of solutions $\lambda$ to
\begin{equation}
        \label{eq-mlx}
\sigma(\lambda)+\varphi+\pi\langle\omega,m\rangle=0\mod \pi,\quad m\in\ZZ^d.
\end{equation}
Each of these eigenvalues
is simple, and for any fixed $m\in\ZZ^d$ Eq.~\eqref{eq-mlx} has
at most one solution $\lambda(m)$, and $\lambda(m)\ne\lambda(m')$
for $m\ne m'$.
\end{prop}

\begin{proof}
As follows from proposition~\ref{prop-krein}, the eigenvalues $\lambda$ of $H$
outside $\spec H^0$ are determined by the condition $\ker\big(M(\lambda)-A\big)\ne 0$,
an their multiplicity coincides with the dimension of the corresponding kernels.
Eq.~\eqref{eq-comm} shows that the condition $(M(\lambda)-A)u=0$
is equivalent to
$\big(1-e^{f_0(\lambda)}\chi U\big)e^{-t(\lambda)}\big(1+\chi U\big)^{-1}u=0$
or, denoting $v:=e^{-t(\lambda)}\big(1+\chi U\big)^{-1}u$,
$(1-e^{f_0(\lambda)}\chi U\big)v=0$, which can be rewritten as
\begin{equation}
    \label{eq-uv}
Uv=\chi^{-1}e^{-f_0(\lambda)}v,\quad v\ne 0.
\end{equation}
As $U$ has the simple eigenvalues $e^{2\pi i \langle \omega,m\rangle}$, $m\in\ZZ^d$,
and the corresponding eigenvectors form a basis, Eq.~\eqref{eq-uv}
implies \eqref{eq-mlx} if one takes into account the identity $f_0(\lambda)=2i\sigma(\lambda)$
proved in lemma~\ref{prop-1}. The rest follows
from the monotonicity of $\sigma$, the inclusion
$\ran \sigma\subset (-\pi/2,\pi/2)$, and the arithmetic properties
\eqref{eq-vf1} and \eqref{eq-vf2}.
\end{proof}

As $[a,b]$ was an arbitrary interval from $\RR\setminus\spec H^0$
and $\spec H^0$ is a discrete set, one has an immediate corollary
\begin{prop}\label{prop-ev2}
The pure point spectrum of $H$ is dense in $\RR$.
\end{prop}

We note that propositions~\ref{prop-ev} and \ref{prop-ev2} automatically imply
$\spec H^0\subset \spec H$ (as $\spec H=\RR$), as $\spec H^0$ is discrete
and lies in the closure of the set of the eigenvalues given by \eqref{eq-mlx}.
We cannot say in general if the Dirichlet eigenvalues are eigenvalues of $H$
and, if it is the case, if they are simple, this depends on
the edge lengths $l_j$ and the edge potentials $U_j$.

\section{Estimates for spectral measures}

Take some $\alpha>0$. For any $\delta>0$ we denote
\begin{equation*}
\SSS^1_\delta=\bigcup_{m\in\ZZ^d} \Big\{
\xi\in \SSS^1: |\Arg \xi-\Arg e^{-2\pi i\langle \omega,m\rangle}|\le \delta\big(1+|m|\big)^{-d-\alpha}
\Big\},\quad
\widetilde\SSS^1_\delta:=\SSS^1\setminus \SSS^1_\delta.
\end{equation*}
Clearly, there holds
\begin{equation}
      \label{eq-ss2}
|1-\xi e^{2\pi i\langle \omega,m\rangle}|\ge 2\pi^{-1}\delta \big(1+|m|\big)^{-d-\alpha},\quad \xi\in \widetilde\SSS^1_\delta,
\quad m\in\ZZ^d.
\end{equation}

Let $\Delta\subset[a,b]$ be an interval whose ends are not eigenvalues of $H$.
Consider the mapping $h:\lambda\mapsto \chi e^{f_0(\lambda)}$.
By~lemma~\ref{prop-1}, $h$ is a diffeomorphism between $\Delta$ and $h(\Delta)$.
By proposition~\ref{prop-ev} one has $h\big(\lambda(m)\big)=e^{-2\pi i \langle\omega,m\rangle}$.
Take an arbitrary $\delta>0$ and denote
\[
\Delta_\delta:=\Delta\cap h^{-1}(\SSS^1_\delta),\quad
\widetilde\Delta_\delta:=\Delta\cap h^{-1}(\widetilde\SSS^1_\delta)\equiv\Delta\setminus\Delta_\delta.
\]
Clearly, $\Delta_\delta$ is a countable union of intervals, and
the limit set $\bigcap_{\delta>0}\Delta_\delta$
coincides with the set of all the eigenvalues $\bigcup_m\{\lambda(m)\}$.

\begin{lemma}\label{lem-fin}
There exists $\varepsilon_0>0$ such that for any $\delta>0$ and any $n\in\ZZ^d$ there exists $C>0$ 
such that
\begin{equation}
  \label{eq-mla}
\big\|\big(M(\lambda+i\varepsilon)-A\big)\theta^n\big\|\le C
\end{equation}
for all $\lambda\in\widetilde\Delta_\delta$,
and $\varepsilon\in(0,\varepsilon_0)$.
\end{lemma}

\begin{proof} Rewrite Eq.~\eqref{eq-MA2} in the form
\[
\big(M(z)-A\big)^{-1}=\big(1+\chi U\big)e^{t(z)}\big(1-e^{f_0(z)}\chi U\big)^{-1}e^{-t(z)} B(z).
\]
Take an arbitrary $n\in\ZZ^d$ and denote $\Psi(z,\theta):=e^{-t(z,\theta)}B(z,\theta)\theta^n$.
Due to the analyticity one can estimate uniformly in $Z$:
\[
|\psi_z(m)|\le C' e^{-\rho|m|},\quad C',\rho>0, \quad \psi_z:=F^{-1}\Psi,\quad
\|(1+\chi U)e^{t(z)}\|\le C'.\nonumber
\]
Therefore, \eqref{eq-mla} follows from the inequality
\begin{equation}
 \label{eq-fini}
\big\|
(1-e^{f_0(\lambda+i\varepsilon)}\chi U)^{-1}\Psi
\big\|\le C.
\end{equation}
Assume that $\varepsilon_0$ satisfies the conditions of lemma~\ref{lem-eps}, then
uniformly for $\lambda\in\Delta$ and $\varepsilon\in(0,\varepsilon_0)$ one has
\begin{multline*}
\big|\big(F^{-1}(1-e^{f_0(\lambda+i\varepsilon)}\chi U)^{-1}\Psi\big)(m)\big|
=\big|
(1-e^{f_0(\lambda+i\varepsilon)}\chi e^{2\pi i \langle\omega,m\rangle})^{-1}\psi_{\lambda+i\varepsilon}(m)
\big|\\
\le 2
\big|
(1-e^{f_0(\lambda)}\chi e^{2\pi i \langle\omega,m\rangle})^{-1}\big| \cdot|\psi_{\lambda+i\varepsilon}(m)|.
\end{multline*}
As in our case $h(\lambda)\equiv\chi e^{f_0(\lambda)}\in \widetilde\SSS^1_\delta$, due to \eqref{eq-ss2}
we have
\[
\big|
(1-e^{f_0(\lambda)}\chi e^{2\pi i \langle\omega,m\rangle})^{-1}\big|\le \dfrac{\pi}{2\delta}\,\big(1+|m|\big)^{d+\alpha}.
\]
Finally,
\begin{multline*}
\big\|
(1-e^{f_0(\lambda+i\varepsilon)}\chi U)^{-1}\Psi
\big\|^2=\sum_{m\in\ZZ^d} \big|\big(F^{-1}(1-e^{f_0(\lambda+i\varepsilon)}\chi U)^{-1}\Psi\big)(m)\big|^2\\
\le \Big(\dfrac{\pi C'}{\delta}\Big)^2\,\sum_{m\in\ZZ^d}\big(1+|m|\big)^{2(d+\alpha)}e^{-2\rho|m|}<\infty,
\end{multline*}
and \eqref{eq-fini} is proved.
\end{proof}

Now we are able to estimate the spectral projections corresponding to $H$.

\begin{lemma}\label{lem-fff} For any $f\in\HH$ and any $\delta>0$ one has
\begin{equation}
  \label{eq-rrr}
\lim_{\varepsilon\to 0+} \varepsilon \int_{\widetilde\Delta_\delta} \|(H-\lambda-i\varepsilon)^{-1}f\|^2d\lambda=0.
\end{equation}
\end{lemma}

\begin{proof}
Here we are going to use proposition~\ref{prop-krein}. First note that due to $\widetilde \Delta_\delta\subset\RR\setminus\spec H^0$ one has
\begin{equation}
     \label{eq-h0}
\lim_{\varepsilon\to 0} \varepsilon \int_{\widetilde\Delta_\delta} \|(H^0-\lambda-i\varepsilon)^{-1}f\|^2d\lambda=0
\text{ for any } f\in\HH.
\end{equation}
Represent $\HH=\HH_0+\HH_1$, where
\[
\HH_0:=\Big(\bigcup_{\Im z \ne 0} \gamma(z)\big(\ell^2(\ZZ^2)\big)\Big)^\perp,
\quad
\HH_1:=\HH_0^\perp;
\]
in other words, $\HH_1$ is the closure of the linear hull of the set $\{\gamma(z)\varphi:\, \Im z\ne 0,\,\varphi\in\ell^2(\ZZ^d)\}$.

By~\eqref{eq-nz}, for any $f\in\HH_0$ and any $z$ with $\Im z\ne 0$ one has $\gamma^*(z)f=0$.
Hence, by~\eqref{eq-krein}, there holds $(H-z)^{-1}f=(H_0-z)^{-1}$, and \eqref{eq-h0}
implies \eqref{eq-rrr} for $f\in\HH_0$.

Now it is sufficient to show \eqref{eq-h0} for vectors $f=\gamma(\zeta)h$
for $h=\big(M(\zeta)-A\big)^{-1}\theta^m$, $m\in\ZZ^d$,
$\Im\zeta\ne0$. The operators $\big(M(\zeta)-A\big)^{-1}$ have dense range
(coinciding with $\dom A$),
hence the linear hull of such vectors $f$ is dense in $\HH_1$.
By elementary calculations (see e.g. section~3 in~\cite{BGP}) one rewrites Eq.~\eqref{eq-krein} as
\begin{equation}
   \label{eq-res}
(H-\lambda-i\varepsilon)^{-1}f=
\dfrac{1}{\zeta-\lambda-i\varepsilon}\,\Big(
f-\gamma(\lambda+i\varepsilon) \big(M(\lambda+i\varepsilon)-A\big)^{-1}\theta^m\Big).
\end{equation}
Due to lemma~\ref{lem-fin} we have $\big\|\big(M(\lambda+i\varepsilon)-A\big)^{-1}\theta^m\big\|\le C$ with some $C>0$,
for all $\lambda\in\widetilde\Delta_\delta$ and sufficiently small $\varepsilon$,
and \eqref{eq-res} implies
\[
\|H-\lambda-i\varepsilon)^{-1}f\|\le \dfrac{\|f\|+C\|\gamma(\lambda+i\varepsilon)\|}{|\zeta-\lambda-i\varepsilon|},
\]
and due to the analyticity of $\gamma$, one can estimate $\|(H-\lambda-i\varepsilon)^{-1}f\|\le C'$
with some $C'>0$ for all $\lambda\in\widetilde\Delta_\delta$ and sufficiently small $\varepsilon$.
This obviously implies \eqref{eq-rrr}.  
\end{proof}

Using the above estimates we are now completing the proof of the main result.

\begin{proof}[Proof of theorem~\ref{th-main}]
The denseness of the pure point spectrum is shown already (proposition~\ref{prop-ev2}).
We are going to show that for any $f\in\HH$ and any interval $\Delta\subset\RR\setminus\spec H^0$
the spectral measure $\mu_f$ associated with $H$ and $f$ satisfies
$\mu_f(\Delta)=\mu_f\big(\Delta\cap\bigcup_m\{\lambda(m)\}\big)$; this proves that all
the spectral measures are pure point.

By the Stone formula, for any set $X$ which is a countable union of intervals
whose ends are not eigenvalues of $H$ one has
\[
\mu_f(X)=\lim_{\varepsilon\to0+}\dfrac{\varepsilon}{\pi}\int_X\|(H-\lambda-i\varepsilon)f\|^2d\lambda.
\]
Using lemma~\ref{lem-fff}, for any $\delta>0$ we estimate
\begin{multline*}
\mu_f(\Delta)=\lim_{\varepsilon\to0+}\dfrac{\varepsilon}{\pi}\int_\Delta\|(H-\lambda-i\varepsilon)f\|^2d\lambda\\
=\lim_{\varepsilon\to0+}\dfrac{\varepsilon}{\pi}\int_{\Delta_\delta}\|(H-\lambda-i\varepsilon)f\|^2d\lambda
+\lim_{\varepsilon\to0+}\dfrac{\varepsilon}{\pi}\int_{\widetilde\Delta_\delta}\|(H-\lambda-i\varepsilon)f\|^2d\lambda\\
=\lim_{\varepsilon\to0+}\dfrac{\varepsilon}{\pi}\int_{\Delta_\delta}\|(H-\lambda-i\varepsilon)f\|^2d\lambda=\mu_f(\Delta_\delta).
\end{multline*}
As $\delta$ is arbitrary and $\bigcap_{\delta>0}\Delta_\delta=\bigcup_m\{\lambda(m)\}$, the theorem is proved.
\end{proof}

\section*{Acknowledgements}

The work was supported by the research fellowship of the Deutsche Forschungsgemeinschaft (PA 1555/1-1),
Sonderforschungsbereich 637 ``Raum, Zeit, Materie'',
the joint German-New Zealand project NZL 05/001 funded by the International Bureau
at the German Aerospace Center, and the DFG-RAS international project 436 RUS 113/785.

\end{document}